\documentclass[epj]{webofc}
\usepackage[varg]{txfonts}   % Web of Conferences font
\usepackage{tikz}
\usetikzlibrary{positioning,decorations.pathreplacing,shapes}
\usetikzlibrary{calc, shapes, backgrounds}
\usetikzlibrary{intersections}
\usetikzlibrary{arrows,decorations.pathmorphing,backgrounds,positioning,fit,petri}
\usetikzlibrary{decorations.markings}
\usetikzlibrary{mindmap}
\usetikzlibrary{calc}
\usepackage{makecell}

\definecolor{colA}{RGB}{241,90,90}
\definecolor{colB}{RGB}{240,196,25}
\definecolor{colC}{RGB}{78,186,111}
\definecolor{colD}{RGB}{45,149,191}
\definecolor{colE}{RGB}{149,91,165}
\DeclareMathOperator{\tr}{tr}

                  \newcommand{\nf}{\ensuremath {n_f}}
                  \newcommand{\nl}{\ensuremath {n_g}}
                  \newcommand{\cR}{\ensuremath {C_F}}
                  \newcommand{\dAA}{\ensuremath {d^{abcd}_Ad^{abcd}_A}}
                  \newcommand{\dFA}{\ensuremath {d^{abcd}_Fd^{abcd}_A}}
                  
		  \newcommand{\NA}{\ensuremath {N_A}}
		  \newcommand{\cA}{\ensuremath {C_A}}
		 \newcommand{\ItoR}{\ensuremath {T_F}}

\woctitle{The 19th International Seminar QUARKS-2016} %
\def\MSbar{\ensuremath{\overline{\mathrm{MS}}}}
\def\ag{\ensuremath{a_s}}
\def\at{\ensuremath{a_t}}
\def\al{\ensuremath{a_\lambda}}
\def\bgG{\ensuremath{\hat G}}
\def\qG{\ensuremath{\tilde G}}
\def\NGen{\ensuremath{n_G}}

\begin{document}
\title{On the four-loop strong coupling beta-function in the SM}

\author{\underline{\firstname{Alexander} \lastname{Bednyakov}}\inst{1}\fnsep\thanks{\email{bednya@theor.jinr.ru}} \and
        \firstname{Andrey} \lastname{Pikelner}\inst{2}\fnsep\thanks{\email{pikelner@theor.jinr.ru}, on leave from BLTP, JINR}
        % etc.
}

\institute{Joint Institute for Nuclear Research,141980 Dubna, Russia
\and
	II. Institut f\"ur Theoretische Physik,  Universit\"at Hamburg, Luruper Chaussee 149, 22761 Hamburg, Germany
          }

\abstract{%
  In the talk the leading four-loop contribution to the beta-function of the strong coupling in the SM is discussed.
  Some details of calculation techniques are provided.  
  Special attention is paid to the ambiguity due to utilized $\gamma_5$ treatment and a particular prescription 
  with anticommuting $\gamma_5$ is advocated.  
  As a by-product of our computation the four-loop beta-function in QCD with ``gluino'' is also obtained.   
  }
\maketitle
%
%\section{Introduction}
%\label{intro}
The Standard Model (SM) of fundamental interactions being renormalizable can, in principle, by used to make predictions at scales far above the 
Z-boson mass $Q^2\gg M_Z$. At such scales it is convenient to use ``running'', or scale-dependent, couplings $a(Q)$, which are obtained from 
a set of measurable quantities $\{O\}$ by means of the following two-step procedure: 
  \tikzset{
    pil/.style={
      ->,
      thick,
      shorten <=2pt,
      shorten >=2pt,}}

\begin{center}
 \begin{tikzpicture}[node distance=0.1cm and 0.5cm, auto,]
        % nodes
	 \node (PDG) {\makecell[c]{\underline{PDG \cite{Agashe:2014kda} 20XX}\\
    $\{O\} = M_b,M_W,M_Z,$\\$\phantom{\{O\}=}M_H,M_t,G_F$}}; 
    \node[right=of PDG] (thresholds)
    {\makecell[c]{\underline{Fixed $\mu_0$}\\
        $g_i(\mu_0),y_i(\mu_0),\lambda(\mu_0)$\\
      in $\MSbar$ scheme}}; 
    \path (PDG.east) edge[pil]
    (thresholds.west); \node[right=of thresholds] (RGE)
    {\makecell[c]{Evolve from $\mu_0$ \\to scale $\mu$}}; 
      \path (thresholds.east) edge[pil]
      (RGE.west);
      \end{tikzpicture}
      \end{center}

The first step 
is called \emph{matching} and boils down to the extraction/fitting 
of the model parameters $a(\mu_0\simeq M_Z)$ at the electroweak scale (in what follows, we employ $\MSbar$-scheme).
The second step --- ``running'' --- allows one to utilize renormalization-group equations (RGEs) to re-summ potentially large logarithms 
$\log \mu^2/\mu_0^2$ contributing to finite-order relations between $a(\mu_0)$ and $a(\mu)$. 

One of the most important applications of such a procedure is the vacuum stability analysis of the SM (see, e.g.,~\cite{Degrassi:2012ry,Bednyakov:2015sca} and references therein). 
It turns out that for large values of Higgs field $\phi$ the effective potential can be approximated as
\begin{align}
	V_\mathrm{eff}(\phi\gg v) \simeq \frac{\textcolor{black}{\lambda}(\mu=\phi)}{4} \phi^4 \label{eq:veff},
\end{align}
	where the scale dependence of  self-coupling $\lambda(\mu)$ is governed by the following (one-loop) RGEs 
\begin{align}
	(4 \pi)^2 \frac{d \lambda}{d \ln \mu^2}  =  12 \lambda + 6 y_t^2 \lambda - 3 \textcolor{colA}{y_t^4} + \ldots, \qquad 
	(4 \pi)^2 \frac{d y_t}{d \ln \mu^2}  = \frac{9}{4} y_t^3 - 4 \textcolor{black}{g_s}^2 y_t +  \ldots, \label{eq:lamdep} 
\end{align}
	in which the ``de-stabilizing'' contribution due to top-quark Yukawa coupling $y_t$ is emphasized.
	The importance of the strong coupling $g_s$ can be deduced from RGE for $y_t$ - strong interactions tend to decrease the latter with $\mu$.

	At present, the state-of-the-art analysis utilizes full two-loop matching \cite{Kniehl:2015nwa} together with three-loop evolution via RGEs \cite{Mihaila:2012fm,Bednyakov:2012en,Chetyrkin:2013wya}.
	In this talk, we discuss one little step towards the full four-loop analysis --- calculation of leading N${}^3$LO corrections to 
	$\beta_{\ag}$. The latter is defined here as ($h$ counts powers of couplings)
\begin{equation}
  \label{eq:betadef}
  \frac{d\;\ag}{d\;\log{\mu^2}}=\beta_{\ag} = - \sum\limits_{i=0}^3\beta_ih^{i+2}.
\end{equation}
	For convenience, we introduce a set of SM parameters (with $\xi$ being a gauge-fixing parameter)
\begin{equation}
	(16 \pi^2) a = \left\{ g_s^2, y_t^2, \lambda, (16 \pi^2) \xi \right\}.
	\label{eq:coupldef}
\end{equation}
	Since we are interested in the leading corrections to $\beta_3$ \eqref{eq:betadef}, 
	the electroweak gauge interactions are neglected together with Yukawa interactions of all SM fermions but the top-quark.  

	For completeness, let us mention here that the \emph{matching} procedure for the strong coupling constant
	is different than that mentioned earlier. One usually considers five-flavor ($n_f=5$) QCD as an effective theory 
	obtained from a more fundamental one (e.g., QCD with ``active'' top quark) and find the relations of the form:
			$$a^{(5)}_s (\mu) = a_s(\mu) \zeta_{a_s}(\mu, M),$$
			where $M$ corresponds to the mass of a heavy field. The (``threshold'') corrections to the so-called decoupling constant $\zeta_{a_s}$ are known
			in pure QCD up to four loops \cite{Chetyrkin:2005ia,Schroder:2005hy,Kniehl:2006bg}, while two-loop electroweak contribution is considered in Ref.~\cite{Bednyakov:2014fua}. 
	
	Before going to the result, let us discuss some technicalities and important issues encountered in our calculation.	
	To simplify our life we made use of the background-field gauge (BFG)~\cite{Abbott:1980hw,Denner:1994xt}.
The advantage of BFG lies in the QED-like relation between the 
gauge coupling renormalization constant $Z_{a_{s}}$ and that of the background gluon field $Z_{\bgG}$:
\begin{equation}
  \label{eq:bgf}
  Z_{a_{s}} = 1/Z_{\bgG},\qquad Z_{\xi} = Z_{\qG}.
\end{equation}
Obviously, this allows one to obtain the final result solely from massless propagator-type integrals. 
In \eqref{eq:bgf}, we also indicate 
the relation between the renormalization constants of quantum gluon field $\qG$ and gauge-fixing parameter.
It is worth mentioning that,  since in $\MSbar$-scheme beta-functions do not depend on masses, one can avoid any special infra-red rearrangement (IRR) \cite{Vladimirov:1979zm} tricks.

For diagram generation we employ the package
\texttt{DIANA}~\cite{Tentyukov:1999is}, which internally
uses~\texttt{QGRAF}~\cite{Nogueira:1991ex}.  
The color \cite{vanRitbergen:1998pn} and Dirac algebra are carried out by means of \texttt{FORM}.
All the generated two-point functions are mapped onto three auxiliary topologies, each containing 11
propagators and 3 irreducible numerators.
The corresponding diagrams are evaluated by means of the \texttt{C++} version of the
\texttt{FIRE} package~\cite{Smirnov:2014hma}, which performs integration-by-parts (IBP) \cite{Chetyrkin:1981qh} reduction 
based on the reduction rules prepared by the \texttt{LiteRed}\cite{Lee:2012cn} package. 
The IBP reduction leads to a small set of master integrals. The expressions for the latter are 
known in analytical form up to the finite parts \cite{Baikov:2010hf}.

Let us also note that as an independent cross-check of our setup, we prepared a simple
QCD-like model with additional fermions in the adjoint representation of SU(3) color group (``gluino'').
We calculated four-loop correction $\Delta \beta_3 \equiv \beta_3(\nf,\nl) - \beta_3(\nf)$ to the beta-function of the strong coupling 
	 \begin{alignat}{2}
		%	&&\beta_3(n_f,n_g)  - \beta_3(n_f)  = \nonumber \\
			\Delta \beta_3/\ag^5 &&= \nl & \biggl[ \frac{\dAA}{\NA}\left(\frac{256}{9} - \frac{832}{3} \zeta_3\right) - 
                    \cA^4 \left(\frac{68507}{243} - \frac{52}{9} \zeta_3\right)\biggr] \nonumber\\
                    %% 
         %           &&+ \nf^2 & \biggl[ 
         %           \cR^2 \ItoR^2 \left(\frac{1352}{27} - \frac{704}{9} \zeta_3\right) + 
         %           \cA^2 \ItoR^2 \left(\frac{7930}{81} + \frac{224}{9} \zeta_3\right) \nonumber\\
         %          &&  & + \cA \cR \ItoR^2 \left(\frac{17152}{243} + \frac{448}{9} \zeta_3\right) +
         %          \frac{\dFF}{\NA} \left(-\frac{704}{9} + \frac{512}{3} \zeta_3\right)\biggr] \nonumber\\
                    %% 
                    && + \nf \nl &\biggl[\cA^2 \cR \ItoR \left(\frac{23480}{243} - \frac{352}{9} \zeta_3\right) + 
                    \cA \cR^2 \ItoR \left(-\frac{152}{27} - \frac{64}{9} \zeta_3\right) + \nonumber\\
                    &&  & \cA^3 \ItoR \left(\frac{30998}{243} + \frac{128}{3} \zeta_3\right) + 
                    \frac{\dFA}{\NA} \left(-\frac{704}{9} + \frac{512}{3} \zeta_3\right)\biggr]\nonumber\\
                    &&+ \nl^2 &\biggl[\cA^4 \left(\frac{26555}{486} - \frac{8}{9}
                      \zeta_3\right) + \frac{\dAA}{\NA} \left(-\frac{176}{9} +
                      \frac{128}{3} \zeta_3\right)\biggr] \nonumber\\
			&&
		    + \nl^2\nf & \biggl[ \cA^3 \ItoR\frac{934}{243} +  \cA^2 \cR \ItoR\frac{308}{243}\biggr] 
		    + \cA^4 \nl^3\frac{23}{27} \nonumber\\
			&& + \nf^2  \nl & \biggl[ \cA^2 \ItoR^2\frac{1252}{243} +  \cA \cR \ItoR^2 \frac{1232}{243}\biggr] 
			%\qquad (\nf/\nl - \mbox{No. of quarks/gluino}) 
		%&& + \nf^2\nl  
                    %% 
			\label{eq:beta_gluino}
		\end{alignat}
			in terms of the  SU(3) casimirs and  $\nf(\nl)$ corresponding to the number of quarks(gluino).  
The beta-function for such a model at four loops was predicted 
by A.F.~Pikelner~\cite{Pikelner:2015} along the lines of Ref.~\cite{Clavelli:1996pz} and can be used, e.g, in the derivation of $\{\beta\}$-expansions \cite{Kataev:2014jba}. 
We found perfect agreement and, thus, both confirmed the prediction and verified our computer setup\footnote{Recently, the result given in Eq.~\eqref{eq:beta_gluino} was also confirmed by an independent calculation \cite{Zoller:2016sgq}.} . 

Let us now discuss an important obstacle -- the ambiguities in the dimensionally regularized expressions due to $\gamma_5$.
It is known that there is a clash between anticommutativity $\{\gamma_\mu, \gamma_5\}=0$ and strictly four-dimensional relation
\begin{equation}
\tr\left( \gamma^\mu \gamma^\nu \gamma^\rho \gamma^\sigma \gamma_5 \right) = -4 i \epsilon^{\mu\nu\rho\sigma}
\label{eq:trace_g5}
\end{equation}
in $D\neq4$ (see, e.g.,\cite{Jegerlehner:2000dz}).
A self-consistent BMHV-algebra \cite{'tHooft:1972fi,Breitenlohner:1977hr} breaks $D$-dimensional Lorentz invariance and requires 
too much effort when applied to multi-loop problems involving chiral fermions. 
External axial currents in QCD can be conveniently treated within the prescription due to Larin \cite{Larin:1993tq}.
Another approach \cite{Korner:1991sx} is based on anticommuting $\gamma_5$ but promote every fermionic trace ``$\tr$'' to a non-cyclic linear functional, which depends 
on the choice of utilized \emph{reading point/prescription}, i.e., the position, at which we start(end) reading the trace. 

\tikzset{
  di/.style={line width=1pt,draw=red, postaction={decorate},
    decoration={markings,mark=at position .55 with
      {\arrow[draw=blue]{>}}}},
  pl/.style={line width=1pt,draw=blue, postaction={decorate},
    decoration={markings,mark=at position .55 with
      {\arrow[draw=black]{>}}}},
  photon/.style={line width=1pt,decorate, decoration={snake}, draw=green},
  gluon/.style={decorate, draw=magenta,
    decoration={coil,amplitude=3pt, segment length=3.5pt}},
  higgs/.style={line width=1pt,draw=blue,dashed}
} 

\begin{figure}[h]
	\centering
	\begin{tabular}{ccc}
		% prescription A
\begin{tikzpicture}[scale=0.7]
	\node[] (bb) at (0,1.3) {};
    \coordinate (c1) at (-0.75,0); \coordinate (c2) at (0.75,0);
    % Left triangle
    \draw[di] (-1,1) -- (c1); % node[midway,below left] {$p_3$};
    \draw[di] (c1) -- (-1,-1);% node[midway,above left] {$p_2$};
    \draw[di] (-1,-1) -- (-2,0);% node[midway,below left] {$p_1$};
    \draw[di] (-2,0) -- (-1,1);% node[midway,above left] {$p_4$};

    % Right triangle
%    \draw[di] (1,1) -- (c2); 
%    \draw[di] (c2) -- (1,-1); 
%    \draw[di] (1,-1) -- (2,0); 
%    \draw[di] (2,0) -- (1,1);
    \draw[di] (1,-1)--(c2); 
    \draw[di] (2,0) -- (1,-1); 
    \draw[di] (1,1) -- (2,0);
    \draw[di] (c2) -- (1,1) node[inner sep=0pt,minimum size=4pt] {}; 
    % Ext legs
    \draw[gluon] (-3,0) -- (-2,0);% node[midway,below=4pt] {$\mu$}; 
    \draw[gluon] (2,0) -- (3,0);
    % Int legs
    \draw (0,1) node[anchor=north] {$\chi/\phi^+$}; 
    \draw[higgs] (-1,1) -- (1,1); 
    \draw ($(c1)!0.5!(c2)$) node[anchor=north] {$h_0/\phi^-$}; \draw[higgs] (c1) -- (c2); 
    \draw[gluon] (-1,-1)  -- (1,-1);% node[midway,below=4pt] {$\rho$};

    % reading points
    % A
    \node[fill=colA,circle,inner sep=0pt,minimum size=5pt] at (-2,0) {}; 
    \node[fill=colA,circle,inner sep=0pt,minimum size=5pt] at (2,0) {}; 

    % B
    %\node<4>[fill=colB,circle,inner sep=0pt,minimum size=5pt] at (-1,1) {}; 
    %\node<4>[fill=colB,circle,inner sep=0pt,minimum size=5pt] at (2,0) {}; 

    % C
    %\node<5,7->[fill=colC,circle,inner sep=0pt,minimum size=5pt] at (-1,1) {}; 
    %\node<5,7->[fill=colC,circle,inner sep=0pt,minimum size=5pt] at (1,1) {}; 

  \end{tikzpicture}
  & 

\begin{tikzpicture}[scale=0.7]
	\node[] (bb) at (0,1.3) {};
    \coordinate (c1) at (-0.75,0); \coordinate (c2) at (0.75,0);
    % Left triangle
    \draw[di] (-1,1) -- (c1); % node[midway,below left] {$p_3$};
    \draw[di] (c1) -- (-1,-1);% node[midway,above left] {$p_2$};
    \draw[di] (-1,-1) -- (-2,0);% node[midway,below left] {$p_1$};
    \draw[di] (-2,0) -- (-1,1);% node[midway,above left] {$p_4$};
    % Right triangle
%    \draw[di] (1,1) -- (c2); 
%    \draw[di] (c2) -- (1,-1); 
%    \draw[di] (1,-1) -- (2,0); 
%    \draw[di] (2,0) -- (1,1);
    \draw[di] (1,-1)--(c2); 
    \draw[di] (2,0) -- (1,-1); 
    \draw[di] (1,1) -- (2,0);
    \draw[di] (c2) -- (1,1) node[inner sep=0pt,minimum size=4pt] {}; 
    % Ext legs
    \draw[gluon] (-3,0) -- (-2,0);% node[midway,below=4pt] {$\mu$}; 
    \draw[gluon] (2,0) -- (3,0);
    % Int legs
    \draw (0,1) node[anchor=north] {$\chi/\phi^+$}; 
    \draw[higgs] (-1,1) -- (1,1); 
    \draw ($(c1)!0.5!(c2)$) node[anchor=north] {$h_0/\phi^-$}; \draw[higgs] (c1) -- (c2); 
    \draw[gluon] (-1,-1)  -- (1,-1);% node[midway,below=4pt] {$\rho$};

    % reading points
    % A
    %\node<3>[fill=colA,circle,inner sep=0pt,minimum size=5pt] at (-2,0) {}; 
    %\node<3>[fill=colA,circle,inner sep=0pt,minimum size=5pt] at (2,0) {}; 

    % B
    \node[fill=colB,circle,inner sep=0pt,minimum size=5pt] at (-1,1) {}; 
    \node[fill=colB,circle,inner sep=0pt,minimum size=5pt] at (2,0) {}; 

    % C
    %\node<5,7->[fill=colC,circle,inner sep=0pt,minimum size=5pt] at (-1,1) {}; 
    %\node<5,7->[fill=colC,circle,inner sep=0pt,minimum size=5pt] at (1,1) {}; 

  \end{tikzpicture}
  &
\begin{tikzpicture}[scale=0.7]
	\node[] (bb) at (0,1.3) {};
    \coordinate (c1) at (-0.75,0); \coordinate (c2) at (0.75,0);
    % Left triangle
    \draw[di] (-1,1) -- (c1); % node[midway,below left] {$p_3$};
    \draw[di] (c1) -- (-1,-1);% node[midway,above left] {$p_2$};
    \draw[di] (-1,-1) -- (-2,0);% node[midway,below left] {$p_1$};
    \draw[di] (-2,0) -- (-1,1);% node[midway,above left] {$p_4$};
    % Right triangle
%    \draw[di] (1,1) -- (c2); 
%    \draw[di] (c2) -- (1,-1); 
%    \draw[di] (1,-1) -- (2,0); 
%    \draw[di] (2,0) -- (1,1);
    \draw[di] (1,-1)--(c2); 
    \draw[di] (2,0) -- (1,-1); 
    \draw[di] (1,1) -- (2,0);
    \draw[di] (c2) -- (1,1) node[inner sep=0pt,minimum size=4pt] {}; 
    % Ext legs
    \draw[gluon] (-3,0) -- (-2,0); %node[midway,below=4pt] {$\mu$}; 
    \draw[gluon] (2,0) -- (3,0);
    % Int legs
    \draw (0,1) node[anchor=north] {$\chi/\phi^+$}; 
    \draw[higgs] (-1,1) -- (1,1); 
    \draw ($(c1)!0.5!(c2)$) node[anchor=north] {$h_0/\phi^-$}; \draw[higgs] (c1) -- (c2); 
    \draw[gluon] (-1,-1)  -- (1,-1);% node[midway,below=4pt] {$\rho$};

    % reading points
    % A
    %\node<3>[fill=colA,circle,inner sep=0pt,minimum size=5pt] at (-2,0) {}; 
    %\node<3>[fill=colA,circle,inner sep=0pt,minimum size=5pt] at (2,0) {}; 

    % B
    %\node<4>[fill=colB,circle,inner sep=0pt,minimum size=5pt] at (-1,1) {}; 
    %\node<4>[fill=colB,circle,inner sep=0pt,minimum size=5pt] at (2,0) {}; 

    % C
    \node[fill=colC,circle,inner sep=0pt,minimum size=5pt] at (-1,1) {}; 
    \node[fill=colC,circle,inner sep=0pt,minimum size=5pt] at (1,1) {}; 

  \end{tikzpicture} \\
$R=1$ & $R=2$ & $R=3$ 
\end{tabular} 
  \centering
  \caption{A typical diagram giving rise to a non-trivial contribution due to traces involving odd number of $\gamma_5$. Three non-equivalent reading prescriptions
  are indicated by dots. In our problem it does not matter, whether we start or end the traces at the indicated points. All internal ``cut'' points turn out to be equivalent.
  }
  \label{fig:readpoints}
\end{figure}
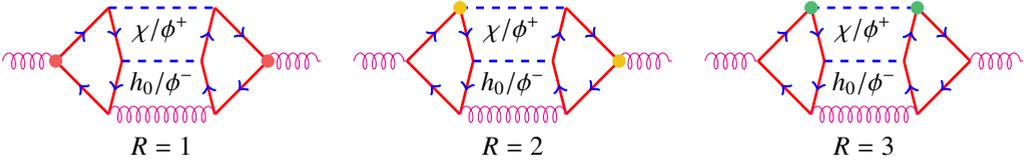

Since the relevant diagrams (48 non-planar and 24 planar graphs, see, e.g., Fig.~\ref{fig:readpoints}) involve only single poles in the regularization parameter 
$\epsilon \equiv (4-D)/2$,  we expected that there should be no ambiguity in $\beta_3$. 
We made a (incorrect)  assumption that it is safe to read a trace from any position and use anticommuting $\gamma_5$, Eq.~\eqref{eq:trace_g5} and the contraction\footnote{Non-trivial contributions 
due to $\gamma_5$ can only appear when even number of such traces are present.}
      \begin{equation*}
        \epsilon^{\mu\nu\rho\sigma} \epsilon_{\alpha\beta\gamma\delta}
        = - {\mathcal T}{}^{[\mu\nu\rho\sigma]}_{[\alpha\beta\gamma\delta]},
        \qquad
        {\mathcal T}{}^{\mu\nu\rho\sigma}_{\alpha\beta\gamma\delta}=
        \delta^\mu_\alpha
        \delta^\nu_\beta
        \delta^\rho_\gamma
        \delta^\sigma_\delta
      \end{equation*}
	to get a unique result.
However, similar calculation was carried out by M. Zoller~\cite{Zoller:2015tha} and an agreement was found only in the ``naive'' part, 
in which contributions due to traces with odd number of $\gamma_5$ are neglected.
The discrepancy triggered further investigation of the issue and it was found that, indeed, the results for the diagrams giving rise to non-trivial $\gamma_5$ contribution 
do depend on the choice of ``cut'' points, at which one breaks a closed Dirac trace.

The result for the $1/\epsilon$ part of the diagrams can be casted into 
\begin{equation*}
\frac{\ag^2 \at^2 T_F^2}{\epsilon} %\left[
\left( X_1 + X_2 \zeta_3 \right) \cdot R
%\right)_{\mathrm{planar}} + \left( X_2 + X_3 \zeta_3 \right)_{\mathrm{non-planar}}\right],
\end{equation*}
		and for non-planar ones we have $X_1 = -1/18$, $X_2 = 1/6$, while in the planar case  $X_1 = 1/6$, $X_2 = 0$.

		The coefficient $R$ depends on the ``cut'' points and it turns out that there are three non-equivalent cases, indicated by dots in Fig.\ref{fig:readpoints}.   
If both traces are cut at external gluon vertices, one has $R=1$.
If only one external vertex is chosen as a ``cut'' point, $R=2$. Finally, for both traces terminated at internal vertices we have $R=3$.   

	A natural question arises whether it is possible to single out a unique prescription. In our original paper \cite{Bednyakov:2015ooa} we advocate the choice $R=3$.  
	The main argument comes from the calculation of finite, $\mathcal{O}(\epsilon^0)$, parts of the diagrams. 
	It is known that IRR procedure (e.g.,of Ref.~\cite{Zoller:2015tha}),  usually utilized to find RGEs in \MSbar{},  is only aimed to calculate the pole part of a diagram and 
does not guarantee that the $\mathcal{O}(\epsilon^0)$ terms remain the same after its application.
	Since we effectively do not do any IRR tricks, we can safely calculate the finite parts and check, whether it is transverse in $D$-dimensions or not\footnote{There seems to be no problem with gauge-invariance in the pole part.}.
	
	It turns out that the case with $R=3$ leads to transverse gluon self-energy, while the case $R=2$ gives rise to a correction to the longitudinal part, thus, explicitly breaking gauge invariance. In spite of the fact that the prescription $R=2$ also produce zero upon multiplication by the product of external momenta $q_\mu q_\nu$, 
we exclude it by simple symmetry argument (we do not want to give preference to either external vertex).  
	
	At the end of the day we obtain the following gauge-parameter independent expression~\cite{Bednyakov:2015ooa}:
\begin{eqnarray}
    \label{eq:beta4sm}
    \beta_3 & = &  \beta_3^{\rm QCD}(n_f = 2 \NGen) 
                  + \ag^4\at \left[ 
                  \ItoR\cR^2\left(   6 - 144 \zeta_3\right ) + 
                  \ItoR\cA\cR\left( \frac{ 523}{9} - 72\zeta_3\right) +
                  \frac{1970}{9}\ItoR\cA^2 \right. \nonumber \\
            & - &
                  \left.\frac{1288}{9}\ItoR^2\cR \NGen -  
                  \frac{872}{9}\ItoR^2\cA \NGen \right]
             + \ag^2 \at^3 \ItoR\left (
                  \frac{423}{2}
                 + 12 \zeta_3
                 \right)
                 + 60 \ag^2 \at^2\al \ItoR
                 - 72 \ag^2 \at\al^2 \ItoR\nonumber\\
                  & - &  \ag^3\at^2 \left[
                  \ItoR^2\left( 48 - 96 \zeta_3 +   
		  \textcolor{colA}{\underbrace{R}_{3}}\cdot \left[
                  \textcolor{colA}{\frac{16}{3}+ 32\zeta_3}
                  \right]
                  \right)+
                  \ItoR\cR\left( 117 - 144\zeta_3\right)+
                  222 \ItoR\cA \right],%\nonumber\\
  \end{eqnarray}	
	where $\NGen$ corresponds to the number of SM families.

\begin{figure}[th]
	     \centering
	    \includegraphics[width=0.60\textwidth]{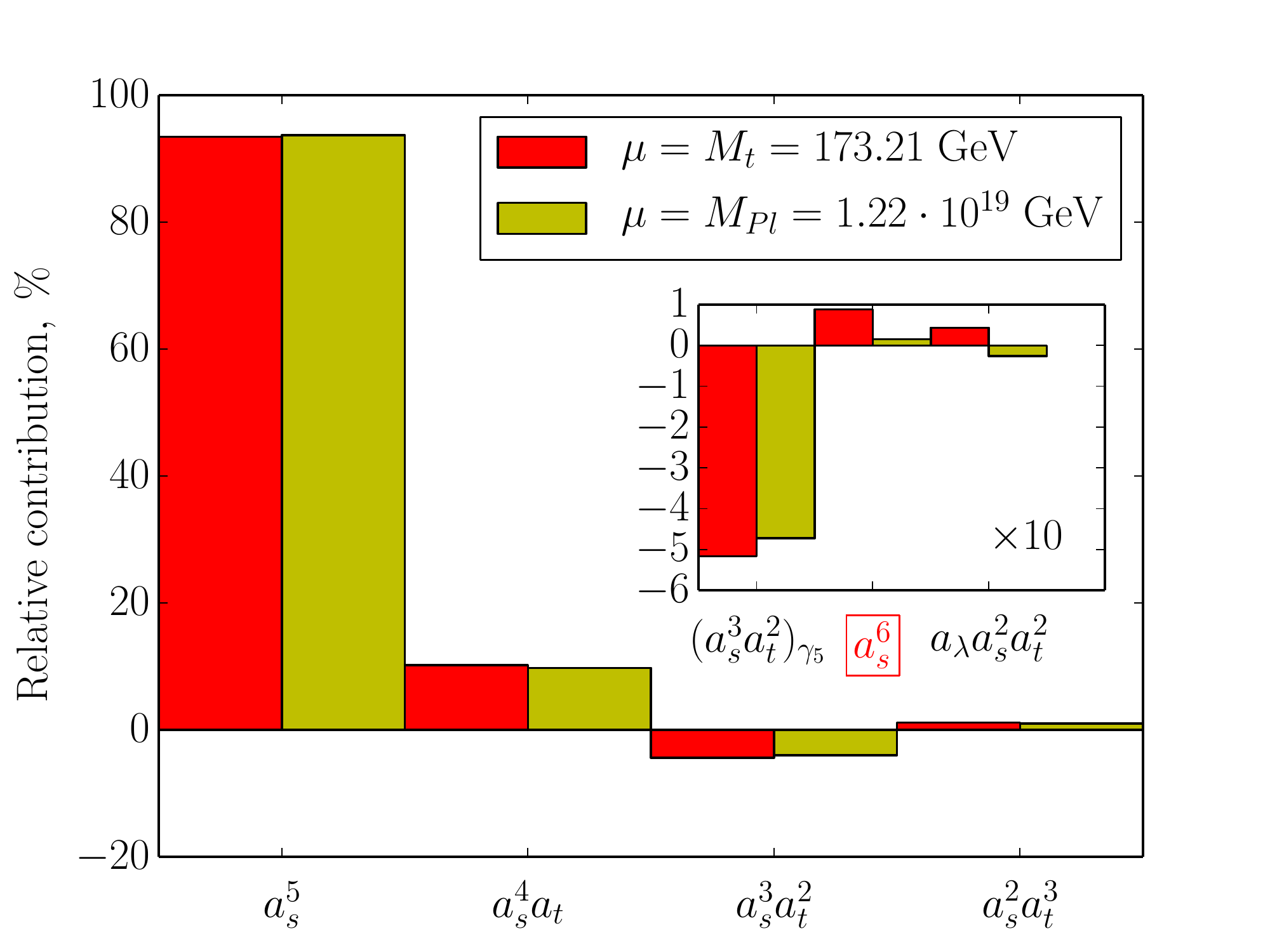}
	    \caption{Relative size of the calculated four-loop
		    contributions and the pure QCD five-loop $\mathcal{O}(\ag^6)$ term 
              with respect to the sum $\beta_3 + \beta_4$. 
	      Non-trivial part due to $\gamma_5$ is indicated. Both the top-mass, $M_t$ \cite{Agashe:2014kda}, and Planck, $M_{Pl}$, scales are considered.  
	     	}
	        \label{fig:rel_c}
\end{figure}

	It is interesting to compare the relative sizes of different four-loop terms \eqref{eq:beta4sm} and recent 
	five-loop pure QCD contribution to $\beta_4$~\cite{Baikov:2016tgj}. 
	From Fig.~\ref{fig:rel_c} one can see that $\ag^5$ amounts for about 94\% of $\beta_3 + \beta_4$  both at the top-mass and Planck scales.
	The mixed $\ag^4 \at$ and $\ag^3 \at^2$ terms have opposite signs and partially compensate each other. 
	The contributions due to  five loops \cite{Baikov:2016tgj} and that from $\gamma_5$ are also of different signs and are both less than a percent. 

	To summarize, we calculated different four-loop corrections to beta-functions for $\alpha_s$ both in the SM and in hypothetical QCD with "gluino". 
	The $\gamma_5$ ambiguities were studied and a reading prescription for "odd" fermion traces, consistent with gauge symmetry, was singled out.  
	In our future studies, we plan to extend the result for $\beta_3$ to the full SM case and compute leading electroweak threshold corrections at three loops.

\section*{Acknowledgments}

	The authors thank the Organizing committee of the QUARKS-2016 seminar for warm hospitality.
	This work is supported in part by RFBR grant 14-02-00494-a and the Heisenberg-Landau programme.

\end{document}